\documentclass[a4paper, 11pt, oneside]{Thesis}  % Use the "Thesis" style, based on the ECS Thesis style by Steve Gunn
\graphicspath{Figures/}  % Location of the graphics files (set up for graphics to be in PDF format)

\usepackage[square, numbers, comma, sort&compress]{natbib}  % Use the "Natbib" style for the references in the Bibliography
\usepackage{verbatim}  % Needed for the "comment" environment to make LaTeX comments
\usepackage{vector}  % Allows "\bvec{}" and "\buvec{}" for "blackboard" style bold vectors in maths

\usepackage{float}
\restylefloat{table}

\DeclareMathOperator*{\argmin}{\arg\!\min}
\DeclareMathOperator*{\argmax}{\arg\!\max}

\begin{document}
\frontmatter      % Begin Roman style (i, ii, iii, iv...) page numbering

% Set up the Title Page
\title  {Automated detection and classification of cryptographic algorithms in binary programs through machine learning}%Automated Detection and Classification of Cryptographic Algorithms in Binary Programs through Machine Learning}
\authors  { Diane Duros Hosfelt }
\addresses  {\groupname\\\deptname\\\univname}  % Do not change this here, instead these must be set in the "Thesis.cls" file, please look through it instead
\date       {\today}
\subject    {}

\maketitle
%% ----------------------------------------------------------------

\setstretch{1.3}  % It is better to have smaller font and larger line spacing than the other way round

% Define the page headers using the FancyHdr package and set up for one-sided printing
\fancyhead{}  % Clears all page headers and footers
\rhead{\thepage}  % Sets the right side header to show the page number
\lhead{}  % Clears the left side page header

\pagestyle{fancy}  % Finally, use the "fancy" page style to implement the FancyHdr headers

%% ----------------------------------------------------------------
% Declaration Page required for the Thesis, your institution may give you a different text to place here
\Declaration{

\addtocontents{toc}{\vspace{1em}}  % Add a gap in the Contents, for aesthetics

I, DIANE DUROS HOSFELT, declare that this thesis titled, `Automated Detection and Classification of Cryptographic Algorithms in Binary Programs through Machine Learning' and the work presented in it are my own. I confirm that:

\begin{itemize} 
\item[\tiny{$\blacksquare$}] This work was done wholly or mainly while in candidature for a research degree at this University.
 
\item[\tiny{$\blacksquare$}] Where any part of this thesis has previously been submitted for a degree or any other qualification at this University or any other institution, this has been clearly stated.
 
\item[\tiny{$\blacksquare$}] Where I have consulted the published work of others, this is always clearly attributed.
 
\item[\tiny{$\blacksquare$}] Where I have quoted from the work of others, the source is always given. With the exception of such quotations, this thesis is entirely my own work.
 
\item[\tiny{$\blacksquare$}] I have acknowledged all main sources of help.
 
\item[\tiny{$\blacksquare$}] Where the thesis is based on work done by myself jointly with others, I have made clear exactly what was done by others and what I have contributed myself.
\\
\end{itemize}

Signed:\\
\rule[1em]{25em}{0.5pt}  % This prints a line for the signature
 
Date:\\
\rule[1em]{25em}{0.5pt}  % This prints a line to write the date
}
\clearpage  % Declaration ended, now start a new page

%%------------------------------------------------------------------

% The Abstract Page
\addtotoc{Abstract}  % Add the "Abstract" page entry to the Contents
\abstract{
\addtocontents{toc}{\vspace{1em}}  % Add a gap in the Contents, for aesthetics

Threats from the internet, particularly malicious software (i.e., malware) often use cryptographic algorithms to disguise their actions and even to take control of a victim's system (as in the case of ransomware).  Malware and other threats proliferate too quickly for the time-consuming traditional methods of binary analysis to be effective.  By automating detection and classification of cryptographic algorithms, we can speed program analysis and more efficiently combat malware.

%	When analyzing a binary program, we are interested in identifying which algorithms are implemented to accomplish the function of the program.  This is especially important when analyzing malicious software (i.e., malware), which often uses cryptographic algorithms to both disguise its actions and possibly take over the victim's computer (as in the case of ransomware).  By detecting and identifying these algorithms automatically, we can make program analysis quicker and easier.
	
This thesis will present several methods of leveraging machine learning to automatically discover and classify cryptographic algorithms in compiled binary programs.

While further work is necessary to fully evaluate these methods on real-world binary programs, the results in this paper suggest that machine learning can be used successfully to detect and identify cryptographic primitives in compiled code.  Currently, these techniques successfully detect and classify cryptographic algorithms in small single-purpose programs, and further work is proposed to apply them to real-world examples.

}

\clearpage  % Abstract ended, start a new page
%% ----------------------------------------------------------------

\setstretch{1.3}  % Reset the line-spacing to 1.3 for body text (if it has changed)

% The Acknowledgements page, for thanking everyone
\acknowledgements{
\addtocontents{toc}{\vspace{1em}}  % Add a gap in the Contents, for aesthetics

Foremost, I'd like to express my appreciation to my advisors, Professor Stephen Checkoway and Professor Matthew Green, for their assistance in shaping and supporting this thesis.  I am deeply grateful to have had the opportunity to learn from them.  Their guidance made this work possible.

I would also like to thank Professor Suchi Saria and Professor Mark Dredze for inspiring my interest in machine learning.  

Overall, the faculty, staff, and students of the JHU Computer Science department have pushed me to learn and think critically, and have motivated me to pursue this Master's degree.  In particular, the members of the JHU Health and Medical Security Lab have encouraged me throughout the past six months and offered their expertise.  I am sincerely thankful to have had the opportunity to work with them all.

Finally, to my friends and family, who let me bounce ideas off of them, even when they had no idea what I was talking about.  Your encouragement has been invaluable.  Most of all, I'd like to thank my cat, Batman, who stood by me (sometimes on the keyboard) throughout this process.

}
\clearpage  % End of the Acknowledgements
%% ----------------------------------------------------------------

\pagestyle{fancy}  %The page style headers have been "empty" all this time, now use the "fancy" headers as defined before to bring them back

%% ----------------------------------------------------------------
\lhead{\emph{Contents}}  % Set the left side page header to "Contents"
\tableofcontents  % Write out the Table of Contents

%% ----------------------------------------------------------------
%\lhead{\emph{List of Figures}}  % Set the left side page header to "List if Figures"
%\listoffigures  % Write out the List of Figures

%% ----------------------------------------------------------------
\lhead{\emph{List of Tables}}  % Set the left side page header to "List of Tables"
\listoftables  % Write out the List of Tables

%% ----------------------------------------------------------------
\setstretch{1.5}  % Set the line spacing to 1.5, this makes the following tables easier to read
\clearpage  % Start a new page
%\lhead{\emph{Abbreviations}}  % Set the left side page header to "Abbreviations"
%\listofsymbols{ll}  % Include a list of Abbreviations (a table of two columns)
%{
% \textbf{Acronym} & \textbf{W}hat (it) \textbf{S}tands \textbf{F}or \\
%\textbf{LAH} & \textbf{L}ist \textbf{A}bbreviations \textbf{H}ere \\

%}

%% ----------------------------------------------------------------
%\clearpage  % Start a new page
%\lhead{\emph{Physical Constants}}  % Set the left side page header to "Physical Constants"
%\listofconstants{lrcl}  % Include a list of Physical Constants (a four column table)
%{
% Constant Name & Symbol & = & Constant Value (with units) \\
%Speed of Light & $c$ & $=$ & $2.997\ 924\ 58\times10^{8}\ \mbox{ms}^{-\mbox{s}}$ (exact)\\

%}

%% ----------------------------------------------------------------
%\clearpage  %Start a new page
%\lhead{\emph{Symbols}}  % Set the left side page header to "Symbols"
%\listofnomenclature{lll}  % Include a list of Symbols (a three column table)
%{
% symbol & name & unit \\
%$a$ & distance & m \\
%$P$ & power & W (Js$^{-1}$) \\
%& & \\ % Gap to separate the Roman symbols from the Greek
%$\omega$ & angular frequency & rads$^{-1}$ \\
%}
%% ----------------------------------------------------------------
% End of the pre-able, contents and lists of things
% Begin the Dedication page

\setstretch{1.3}  % Return the line spacing back to 1.3

\pagestyle{empty}  % Page style needs to be empty for this page
\dedicatory{Dedicated to the colleagues, professors, friends, family, and cat who made this possible}

\addtocontents{toc}{\vspace{2em}}  % Add a gap in the Contents, for aesthetics

%% ----------------------------------------------------------------
\mainmatter	  % Begin normal, numeric (1,2,3...) page numbering
\pagestyle{fancy}  % Return the page headers back to the "fancy" style

% Include the chapters of the thesis, as separate files
% Just uncomment the lines as you write the chapters

%headers are off by a page, just not dealing with it
\lhead{}%\emph{Introduction}}
\chapter{Introduction}

Detection and mitigation of malicious software (malware) is an evolving problem in the security field.  As researchers develop new techniques, malware authors improve their ability to evade detection.  One way they do this is by leveraging cryptovirology, where they use cryptography to disguise certain activities \cite{cryptovirology}.

Like any other class of malware, cryptoviruses can take many forms, such as malware that logs user activity or installs malicious software.  Recently, ransomware, where an attacker holds a system captive and demands a ransom, has become prevalent.  Recent viruses, such as CryptoLocker, encrypt the user's files, then direct them to pay a ransom via Bitcoin.  These viruses rely heavily on cryptography for their attacks.

CryptoLocker was discovered in September 2013, and by the time a solution was delivered to the public in August 2014, the virus had affected approximately 500,000 computers and extorted around \$3 million \cite{BBCcryptolocker}.  For encryption, CryptoLocker used the public key encryption scheme RSA with a 2048-bit key pair.  Once the ransom had been paid, the user received the private key for decryption (although this did not always occur).  International law enforcement agencies disrupted the command-and-control servers and intercepted the database of private keys, which was made available to users in August 2014 \cite{ARScryptolocker}.  

Since CryptoLocker, ransomware has become increasingly popular.  Other recent crypto viruses are:
\begin{itemize}
	\item CryptoWall: Andrea Allievi and Earl Carter from Cisco's Talos group analyzed CryptoWall 2.0.  It uses The Onion Router (TOR) to protect communication with the command and control channel and encrypts files using 2048-bit RSA \cite{cryptowall}.
	\item TorrentLocker: iSIGHT Partners has analyzed this ransomware through a number of iterations.  It uses AES (Rijndael) encryption, with either 128, 192, or 256-bit keys, and initially encrypted all files with the same key using Output Feedback (OFB) mode. After this disclosure, the authors modified it to use Cipher-Block Chaining (CBC) mode with different key streams \cite{torrentlocker}. 
	\item Critroni/Onion: Kaspersky Labs analyzed Onion and discovered that it compresses files then encrypts them using AES with the hash SHA256.  Decryption is only possibly by using Elliptic Curve Diffie-Hellman (ECDH) with a master-private key, stored on the authors' server.  All communication with this server is also protected by ECDH with a different set of keys via TOR \cite{onion}. 
\end{itemize}

Due to the prevalence of cryptographic primitives in malware, researchers are interested in efficiently reverse-engineering program binaries to detect crypto.  This requires labor-intensive manual analysis of the binaries, which does not scale well to the number of emerging threats.  

The goal of this thesis is to utilize machine learning to detect and classify cryptography in small (single purpose) programs.  This technique may then be extended to larger, real-world examples by isolating the basic blocks (defined in chapter 2) of a program, and applying it against each basic block.  Previous work in this area relies on heuristics \cite{grobert}\cite{caballero:dispatcher}.  Unlike heuristics, it is easy to update machine learning models as malware authors evolve their evasion capabilities, and they will be more robust to obfuscation attempts.  Instead of constantly tweaking a heuristic or threshold based on new malware samples, models can be re-trained with new data, which will scale better as malware continues to proliferate.

The primary contributions of this thesis are:
\begin{itemize}
	\item Providing a framework for generating machine learning models that detect and classify cryptographic algorithms in binary programs
	\item Demonstrating that these models are successful when tested against single-purpose binary programs
	\item Suggesting further work that will build upon the models presented and extend them to successfully detect and classify crypto in larger (multipurpose) binary programs.
\end{itemize}

In chapter 2, I will discuss related work and how it has formed the foundation for this thesis.  Chapter 3 details the methodology I used to extract features from binary programs in order to perform machine learning, which is then explored in chapter 4.  Finally, I evaluate the results of my experiments in chapter 5, and examine the limitations in chapter 6, as well as proposing future work, before concluding in chapter 7.   % Introduction

%\fancyhead{}
\lhead{}%\emph{Introduction}}
\chapter{Related Work}

Much of the previous work in this area builds on the field of automated protocol reverse-engineering [\cite{caballero:polyglot},\cite{lin},\cite{Wondracek}], where the application-level protocol is extracted without access to the specification itself.  Like the task of analyzing binary programs for cryptographic primitives, protocol reverse-engineering is a time consuming manual task.  In general, the previous work has focused on creating domain-specific heuristics, instead of applying general machine learning algorithms; however, machine learning has been used successfully to detect and classify malware.

\section{Automated protocol reverse engineering}
Previous work in this area has been split between static approaches and dynamic approaches.  Static tools rely on signatures to determine if a particular implementation of a cryptographic primitive is present in a binary sample.  Wang et al. \cite{wang} were one of the first groups to implement a dynamic approach using data lifetime analysis and dynamic binary instrumentation (DBI) to identify when message decryption and processing occur, and extract the decrypted message from memory.

Caballero et al. \cite{caballero:dispatcher} refined this technique to automatically reverse engineer the MegaD malware, which uses a custom encryption protocol to evade detection. The authors identified a number of heuristics concerning loop detection and the ratio of bitwise arithmetic instructions.  In this case, they required 20 executions of a function that had a ratio of at least 55\% bitwise arithmetic instructions.

Lutz \cite{lutz}, cited by Caballero et al\cite{caballero:dispatcher}., created a tool that automatically decrypts encrypted input received by malicious binaries, using dynamic analysis to extract features.  This approach searches through the feature space to discover the location of the decryption routine and the decrypted input in system memory.  He also noted the two features of cryptographic code that this thesis (and Cabellero et al.) rely upon:
\begin{enumerate}
	\item Cryptographic code uses a high ratio of bitwise arithmetic instructions
	\item Cryptographic code contains loops (but they are an insufficient feature on their own)
\end{enumerate}

\section{Classifying malware with machine learning}
Kolter and Maloof\cite{kolter} used machine learning to detect and classify malware executables.  They collected 1971 benign executables and 1651 malicious executables for the detection task, as well as various virus loaders, worms, and Trojan horses for classification, then converted each binary executable to a hexadecimal representation.  For features, they extracted $n$-grams from the hexadecimal strings, and approached the classification task as a text classification problem.  

They evaluated the performance of support vector machines (SVMs), naive Bayes, decision trees, and boosted classifiers, and determined that a boosted decision tree performed the best in both the detection and classification tasks.

\section{Detecting cryptographic primitives}

Gr\"{o}bert et al. \cite{grobert} created three heuristics to detect cryptographic basic blocks: chains, mnemonic-const, and verifier.  The chains heuristic is based on the ordering of instruction sequences.  The authors compiled a database of known instruction sequences generated from different open-source crypto implementations, which they then compared to sequences found in an unknown sample.

The mnemonic-const heuristic combines the chains heuristic with an examination of constants.  They noted that each algorithm contains unique tuples, which they consider characteristic for the algorithm.  They check for patterns using the chains heuristic, then intersect the results with those of the characteristic tuples.

The verifier heuristic confirms a relationship between the input and output of a permutation box using memory reconstruction.  This allows the authors to reconstruct candidates for the key, plaintext, and ciphertext, which they use to verify the implementation of the algorithm and extract its parameters.

Like Caballero et al.\cite{caballero:dispatcher}, Gr\"{o}bert et al use specific heuristics, with empirically determined thresholds to perform their detection classification.  

The authors also focus on the importance of basic block detection for successful detection, where a basic block is a sequence of instructions in a given order that has a single entry and exit point.  These are generated from the dynamic trace.  Utilizing basic block detection is suggested as further work for this thesis.

While Lutz, Lin et al., and Wang et al. utilized Valgrind to perform DBI, Gr\"{o}bert et al. used Intel's Pin\cite{pin} DBI.  As the following work most directly relates to that of Gr\"{o}bert et al., we elected to use used Pin, which is described in further detail in the next section.  This work extends the previous work by utilizing machine learning instead of heuristics for detection and classification.

 % Background Theory 

%\fancyhead{}
\lhead{}%\emph{Feature Extraction}}
\chapter{Feature Extraction}

\section{Data}
To generate data, small programs were created using open source code with the sole function of implementing certain cryptographic algorithms, using the OpenSSL implementation for each algorithm.  I trained my models on data generated from the following cryptographic algorithms:
\begin{enumerate}
	\item AES (CBC mode)
	\item RC4
	\item RSA
	\item Sha1
	\item MD5
	\item Triple DES (CBC, CFB, OFB modes)
\end{enumerate}
 For encryption algorithms, the programs implemented encryption and decryption.  For the hashing algorithms, two programs were created: one that simply called the hash function and one that called HMAC implemented with the specific function.

In order to generate enough training data, I then compiled each small program with different compiler flags and compilers (gcc and clang), along with some programs that did not contain crypto.  This resulted in 317 binary files.

\section{PIN framework}

The first task in applying machine learning to this problem is to identify and extract features.  I used Intel's Pin dynamic binary instrumentation (DBI) framework \cite{pin} to generate these features.  DBI enables an analyst to examine the behavior of a binary program at runtime by injecting instrumentation code.  As the code executes, DBI tools analyze what actually occurs, instead of considering what might occur (as in static binary analysis).

To perform instrumentation, Pin just-in-time (JIT) recompiles fragments of the binary executable immediately before running it, then inserts the instrumentation code.

Lutz \cite{lutz} noted that cryptographic algorithms have uncommonly high counts of bitwise arithmetic instructions and loop executions.  Three types of features were generated: instruction, category, and loop.  For the instruction and category features, models were created that used the proportion of each feature to the total number of instructions in the program, as well as the raw count.  Unlike Gr\"{o}bert et al.\cite{grobert}, here we focus solely on dynamic features, and do not consider any static features.

\subsection{Instruction features}
It is a simple task to use Pin to track the number of times an individual instruction is executed.  Each value in an instruction feature vector corresponds to either the count or the proportion of each individual instruction to the total number of instructions.  %The index is given by the opcode output by PIN.

\subsection{Category features}

Pin classifies instructions into categories, such as NOP, SYSCALL, BINARY, STRINGOP.  For this feature vector, instead of tracking the counts of individual instructions, we tracked the instruction categories.  Since cryptographic algorithms have disproportionately high counts of arithmetic instructions, the goal for this feature set was to capture this at a higher level than examining individual instruction counts.

 %Each category is indicated by an id, and the ids produced by my training examples ranged from 1 to 60.  

\subsection{Loop features}
I utilized an open source Pin tool to detect loops using the instruction counter \cite{simpleloop}, and recorded the number of times the loop was executed.  Initial testing revealed that there were only 8 instructions that were repeatedly looped over in the crypto examples
\footnote{One additional loop feature that could be of interest would be bigrams of instructions, that is, seeing what instructions are executed sequentially.  It could also be interesting to consider a simple cross product of instruction/category features and loop features}
:
\begin{enumerate}
	\item push qword ptr [addr]
	\item jmp qword ptr [addr]
	\item repne scasb byte ptr [reg]
	\item add x, y
	\item lea x, y
	\item not x
	\item and x, y
	\item jz addr	
\end{enumerate}

%To create features, I summed the number of times each was called in a loop, resulting in a vector of length 8.

\section{Feature extraction results}
When training the models, either the instruction features or the category features were used to avoid redundancy.  Any information learned from the individual instructions should be encapsulated in the category features, so we then compared the models generated from each type to determine whether the high granularity of the instruction features was justified given the larger (19x) vector size.

Two sets of each feature type were also generated, one which stored the raw count per instruction/category, and one which stored the proportion of the instruction count to all of the instructions.  Intuitively, the proportional features should be more informative than raw counts.

%To compare the granularity of instruction-based features versus the higher level category features, I extracted two sets of features.  %The first is a vector of length 1142 that encapsulates 1138 instructions and 8 loop counts.  The second is a vector of length 69 that encapsulates 60 categories and 8 loop counts.  For simplicity, and to maintain consistency with Pin's ids, the zero index of both the category and instruction vectors holds the value 0.

\subsection{Timing Analysis}
Due to the small size of the training set as well as the sparse nature of the feature vectors, model training was fast (less than 1 second per model).  Feature extraction required significantly more time.  The average time to process features was slightly over 3 seconds.  The feature extraction processing was evaluated on 169 training files.  There were some failures where Pin failed to extract loop features.

\begin{center}
\begin{table}[H]
\begin{tabular}{c|cc}
\textbf{Feature type} & \textbf{Average processing time (seconds)} & \textbf{Total failures}\\
\hline
Instruction & 1.779 & 0\\
Category & 1.444 & 0\\
Loop & 1.654 & 59
\end{tabular}
\label{featureprocessing}
\caption{Summary of processing failures and timing analysis for feature extraction}
\end{table}
\end{center}
 % Experimental Setup

\lhead{}%\emph{Machine Learning}}
\chapter{Machine Learning}

\section{Models}

Three types of models were implemented:
\begin{enumerate}
	\item Detection model
	\item Type classification model
	\item Algorithm classification model
\end{enumerate}

The first model classified examples as having crypto versus not having crypto.  The second determined if the crypto present is encryption or hashing, and the third tried to classify it as a particular algorithm.  For evaluation, I implemented each model under all four learning algorithms detailed below.

\section{Framework}

To perform machine learning on extracted features, I utilized the scikit-learn \cite{scikit-learn} package for Python.

\section{Supervised Learning}
\subsection{Support Vector Machine (SVM)}
A linear SVM for binary classification processes labeled training data in order to define a hyperplane that separates the two classes.  While there are many hyperplanes that could separate the data, the best is accepted to be the hyperplane that creates the largest separation between the two classes, or the maximum margin.  This relies on the assumption that the data are linearly separable.  Once the SVM is trained, the model stores a weight vector $\lambda$ such that an unbiased hyperplane takes the form $\lambda^T x = 0$. 

The classifiers will take the form
\begin{equation}
	\hat{y} = \sum_{i=1}^n \lambda_i K(x, x_i) \label{svm}
\end{equation}

In order to properly represent possibly nonlinear data, we can use a linear SVM with a kernel function, $K(x, x')$.  We transform the data $x$ with some basis function $\phi(x)$, which projects the data into a higher dimensional space, where it will be separability can be achieved.  This allows nonlinear classification in a parametric linear framework.

The detection and type classification models are both binary classification problems, while the algorithm classification model is a multi-class problem.  For this, we used scikit-learn's SVC class, which implements a one-versus-one model \cite{knerr}, and results in $n (n-1)/2$ pairwise binary classifiers for $n$ classes.

\subsection{Kernels}
Scikit-learn supports 4 different kernel functions: radial-basis functions (RBF), linear, polynomial, and sigmoid functions.  The default parameters provided by scikit-learn were used, so improved performance could potentially be achieved by optimizing them.%I used the default parameters, so it is likely there could be significant improvement given parameter tweaking.
 
 %reference: http://crsouza.blogspot.com/2010/03/kernel-functions-for-machine-learning.html
\begin{description}
	\item[RBF]: This is a Gaussian kernel.  Given two samples $x, y$
		$$ K(x, y) = \exp \left( -\frac{\mid \mid x-y \mid \mid^2}{2\sigma^2} \right) $$
	\item[Linear]: This is the simplest kernel function
		$$ K(x, y) = x^Ty + c $$
	\item[Polynomial]: The default degree for this kernel is 3.  Given two samples $x, y$,
		$$ K(x, y) = (\alpha x^Ty +c)^d $$ %large degrees tend to overfit
	\item[Sigmoid]: This kernel originated from the neural networks field, and performs well in practice
		$$ K(x, y)= \tanh(\alpha x^T y +c) $$
\end{description}

\subsection{Naive Bayes}
In a naive Bayes model, we assume that all features are conditionally independent from each other.  Despite this strong assumption, in practice, naive Bayes classifiers perform very well, particularly in document classification problems.

Two types of naive Bayes classifiers were evaluated: Gaussian and multinomial.  The Gaussian implementation assumes that the continuous values adhere to a Gaussian distribution, and given the mean and variance of the feature vectors in each class $c$, we compute the probability distribution of a feature vector given a class.

\begin{equation}
	p(x= v \mid c) = \frac{1}{\sqrt{2\pi\sigma_c^c}}e^{-\frac{(v-\mu_c)^2}{2\sigma_c^2}}\label{Gaussianbayes}
\end{equation}

The multinomial model is often used in document classification.  Features in the current work were considered to be similar to those in a text document, where data are either represented as word vector counts or tf-idf features (term frequency and inverse document frequency).  In this scenario, the number of times an instruction was executed was considered analogous to a word count.  Despite this, multinomial naive Bayes performed significantly worse than the Gaussian variant (detailed in chapter 5).  The likelihood of observing a feature vector $x$ in class $C_k$,

\begin{equation}
	p(x \mid C_k ) = \frac{(\sum_i x_i)!}{\prod_i x_i !}\prod_i p_{ki}^{x_i}
	\label{multinomialbayes}
\end{equation}

To assign a classification label, the model picks the class that maximizes the probability that a testing instance $x_i$ belongs to class $C_k$

\begin{equation}
\hat{y} = \argmax_{k\in\{1\ldots K\}} p(C_k) \prod_{i=1}^n p(x_i \mid C_k) \label{naivebayes}
\end{equation}

\subsection{Decision Tree}
As opposed to naive Bayes and SVMs, a decision tree implements non-parametric learning.  Instead of training certain parameters, the model learns decision rules from the training features.

Each internal node is a decision rule, and the leaf nodes are classes.  By filtering a feature pattern through the nodes, the model reaches a classification decision.

\section{Unsupervised Learning}
\subsection{K-means Clustering}
Given a predefined value $k$, this algorithm partitions the data into $k$ clusters without the use of labels by minimizing the within-cluster sum of squares (Euclidean distance):

\begin{equation}
\argmin_C \sum_{i=1}^k \sum_{x\in C_i} \mid \mid  x-\mu_i \mid \mid ^2\label{kmeans}
\end{equation}

To evaluate the accuracy of this method, the clustered data is compared to the labeled data to determine if instances were clustered appropriately.

Given a dataset with well-defined classes, the choice of $k$ was straightforward.  The first two models were characterized in terms of a binary classification problem, so $k = 2$.   For the third task, the number of clusters depends on the number of suspected algorithms in the evaluation set. Given a priori knowledge that the current set contained 6 algorithms, the k-means algorithm was set to create 6 clusters. In practice, the ÒtrueÓ or optimal value of k is not easily determined.

%For the first two models, we are considering a binary classification problem, so $k=2$.  For the third task, the number of clusters will depend on the number of suspected algorithms in the evaluation  set.  Due to a priori knowledge of the test set, we can instruct the model to generate 6 clusters.  In practice, the value of $k$ is not so easily determined.  %For the third model, I am only training on 6 algorithms, so $k=6$. 

%\textbf{dumb thats just not how this works}%%%

Scikit-learn uses Lloyd's algorithm \cite{lloyd} to solve the k-means problem.

 % Experiment 1

\lhead{}%\emph{Experimental Evaluation}}
\chapter{Experimental Evaluation}
All experiments were run on an Ubuntu virtual machine with 1 processor core and 1024 MB of memory.

\section{Measures}
\subsection{Supervised learning}
The supervised algorithms were evaluated in two ways: cross-validation within the original dataset and validation on a separate dataset.

\textbf{Cross-validation}: Three-fold cross-validation was used to evaluate each model; 3 folds were used instead of the more common 10 due to the small size of the dataset.  The training data were split into 3 equally sized partitions, with 2 subsets used to train the model, and the remaining subset used for validation.  This was repeated 3 times and producesd 3 different accuracy scores, which were then averaged.

%\textbf{See what happens when you use StratifiedKFold}

\textbf{Validation on separate data}:  In addition to using cross-validation on the original training set for evaluation, we used separate testing data.  Initially, the models were tested against real-world, multipurpose programs; however, they performed poorly, due to the difference in scope between the training and testing data.  Then, the models were tested against the same small, single-purpose programs that had been compiled with a separate compiler.

We believe that this testing dataset was too similar to the training data, leading to overly optimistic results on the testing data.

 %However, we believe that the data sets are too similar, leading to overly optimistic results on the testing data.  Due to this, we expect that the cross-validation results are more realistic.

To test Model 1, open source implementations of various crypto algorithms were aggregated, as well as programs that do not contain crypto.

To test Models 2 and 3, the training set was compiled first with gcc, then with clang.  Each program in the dataset is single-purpose and only implements the cryptographic algorithm to emulate the idea of a basic block and isolate the crypto.  %When these models were tested against larger programs, they performed poorly without this isolation.

After fitting the model, three measures of correctness were evaluated: precision, recall, and f1-score (averaged over all classes).  Consider the true positives ($tp$), true negatives ($tn$), false positives ($fp$), and false negatives ($fn$).
\begin{description}
\item[Precision]: High precision indicates that the model returns more relevant results than irrelevant $$\frac{tp}{(tp+fp)}$$
\item[Recall]: High recall indicates that the model returns most of the relevant results $$\frac{tp}{(tp+fn)}$$
\item[F1-score]: The harmonic mean of the precision and real $$f1=2\times(\text{precision}\times\text{recall})/ (\text{precision}+\text{recall})$$
\end{description}

\subsection{Unsupervised learning}
To evaluate the models generated by k-means clustering, three metrics were applied: homogeneity, completeness, and v-score \footnote{Other metrics such as the Silhouette Score of the stability analysis in Lange et al. \cite{lange} would be more appropriate in scenarios when the true $k$ is not known.}.  Given true class assignments of samples, these metrics measure how well the data were partitioned.

\begin{description}
	\item[Homogeneity]: A cluster is homogeneous (value = 1.0) if all clusters contain data points which belong to a single class
	\item[Completeness]: A cluster is complete (value = 1.0) if all data points of a class are elements of the same cluster
	\item[V-score]: The harmonic mean between homogeneity and completeness $$v=2\times(\text{homogeneity}\times\text{completeness})/ (\text{homogeneity}+\text{completeness})$$
\end{description}

\section{Experimental Results}
\subsection{SVM kernel Results}
To determine the optimal kernel function, 3-fold cross-validation was performed on a dataset that consisted of 317 total training files.  These were compiled with two different compilers (gcc and clang) and a variety of compiler options.  The default parameter values assigned by scikit-learn were used.  As shown in Table \ref{table:kernelfunction}, the linear kernel has the best overall performance.  

The remainder of the results are derived from the SVM with a linear kernel.

\begin{center}
\begin{table}[H]
\begin{tabular}{c|cccc}
\textbf{Kernel} & \textbf{Model 1} & \textbf{Model 2} & \textbf{Model 3} & \textbf{Average}\\
\hline
Linear & .98 &.94 & .92 & .95\\
RBF & .96 & .90 & .76 & .87\\
Polynomial & .98 & .94 & .84 & .92\\
Sigmoid & .96 & .59 & .29 & .62
\end{tabular}
\caption{The accuracy of each SVM kernel function on each model}
\label{table:kernelfunction}
\end{table}
\end{center}

\subsection{Feature type results}
Each model was evaluated on 4 feature types (category vs. instruction and count vs. proportion) to determine the optimal feature set, as seen in Table \ref{table:feature}.  In the clustering task, the counting features performed nearly twice as well as the proportional features, while proportional features performed better on the classification task.  There was no significant difference in the overall performance between instruction and category features.  This is not unexpected, as the category feature would encapsulate the instruction feature.

For the remainder of the results, we will focus on results from the proportional and categorical feature set for the classification task, and the counting and instruction feature set for the clustering task.

\begin{center}
\begin{table}
\begin{tabular}{c|cccc}
& \textbf{Count/ins} & \textbf{Count/cat} & \textbf{Prop/ins} & \textbf{Prop/cat}\\
\hline
\textbf{Avg. cross-validation} & .939 & .927 & .987 & .984\\
\textbf{Avg. f1-score} & .860 & .837 & .961 & .972\\
\textbf{Avg. v-score} & .460 & .460 & .253 & .255\\

\end{tabular}
\caption{Summary of feature set performance as measured by average cross-validation score, f1-score, and v-score}
\label{table:feature}
\end{table}
\end{center}

\subsection{Model 1: Cryptographic detection binary classifier}
The results for detecting cryptographic algorithms with each different machine learning algorithm are shown in Table \ref{table:model1}.  It is possible that the testing data were too similar to the training data, and not general enough, resulting in biased results.

\begin{center}
\begin{table}
\begin{tabular}{c|ccc}
\textbf{Algorithm} & \textbf{Precision/Homogeneity} & \textbf{Recall/Completeness} &\textbf{F1-score/V-score}\\
\hline
SVM & 1 & 1 &1\\
Naive Bayes & 1 & .97 &.99\\
Decision Tree & 1 & 1 & 1\\
K-means & 1 & 1 & 1\\
\end{tabular}
\caption{Summary of classification/clustering results for a model that classifies a binary program as containing crypto or not (Model 1)}\label{table:model1}
\end{table}
\end{center}

\subsection{Model 2: Cryptographic algorithm type binary classifier}
The results for classifying algorithms as encryption or hashing with each different machine learning algorithm are shown in Table \ref{table:model2}.

\begin{center}
\begin{table}
\begin{tabular}{c|ccc}
\textbf{Algorithm} & \textbf{Precision/Homogeneity} & \textbf{Recall/Completeness} &\textbf{F1-score/V-score}\\
\hline
SVM & 1 & 1& 1\\
Naive Bayes & .92 & .9& .9\\
Decision Tree & 1 & 1& 1\\
K-means & 0 & 1 & 0\\
\end{tabular}
\caption{Summary of classification/clustering results for a model that classifies a binary program that contains crypto as having encryption or hashing (Model 2)}\label{table:model2}
\end{table}
\end{center}

\subsection{Model 3: Cryptographic algorithm multiclass classifier}
The results for classifying each cryptographic algorithm with each different machine learning algorithm are shown in Table \ref{table:model3}.
\begin{center}
\begin{table}
\begin{tabular}{c|ccc}
\textbf{Algorithm} & \textbf{Precision/Homogeneity} & \textbf{Recall/Completeness} &\textbf{F1-score/V-score}\\
\hline
SVM & 1 & 1& 1\\
Naive Bayes & .83 & .9& .86\\
Decision Tree & 1 & 1& 1\\
K-means & .26 & .71& .38 \\
\end{tabular}
\caption{Summary of classification/clustering results for a model that classifies a binary program that contains crypto as belonging to one of five pre-specified algorithms (Model 3)}\label{table:model3}
\end{table}
\end{center}

\subsection{Discussion}
These results demonstrate that the best SVM kernel function for this problem is a linear kernel, and that while proportional and categorical features work best for the classification task, it is better to use counting and instructional features for the clustering task.  As shown in Table \ref{table:allcv}, the decision tree models perform the best in cross-validation testing.

It is likely that the testing data are too similar to the training data, implying that the models work perfectly.  In this case, we can utilize on the results of cross-validation over the entire training set (using both gcc and clang for compilation) to conclude that the models will likely generalize given test data that are less similar to the training data.  Additionally, Model 1 would benefit from training on more non-cryptographic examples, due to bias from cross-validating with unequal numbers of samples per class.

When the models were evaluated on larger real-world programs, they performed very poorly, likely due to the difference in scale of the training and testing examples.  While the current models are trained on very small programs that perform a specific task, real-world examples will be large programs that perform many tasks. In order to succeed on these real-world examples, further work is needed.

\begin{center}
\begin{table}
\begin{tabular}{c|ccc}
&\textbf{Model 1} & \textbf{Model 2} &\textbf{Model 3}\\
\hline
SVM & .984 & .944 & .927\\
Naive Bayes & .741 & .897 & .779\\
Decision Tree & .984 & .993 & .987
\end{tabular}
\caption{cross-validation results for each model and classification algorithm}
\label{table:allcv}
\end{table}
\end{center}

 % Experiment 2

\lhead{}%\emph{Limitations}}
\chapter{Limitations}

This method relies on dynamic analysis using Pin for feature extraction.  If the code of interest is not executed during instrumentation, then it will not be analyzed and extracted.  Therefore, we must assume that the cryptographic code is always executed.  However, if malware can detect the presence of the Pin instrumentation code, it would be able to avoid executing this code and thus avoid detection. As mentioned in \cite{grobert}, using a more robust malware analysis framework for feature extraction could solve this.

In this thesis, only C/C++ compiled code has been examined.  If an attacker uses a language such as Python, analysis would become complicated.  However, once appropriate features are identified and extracted, it would not be difficult to train models to detect and classify crypto in Python binaries.

\section{Model limitations}
As discussed in the previous chapter, the results suggest that a decision tree is the best learning algorithm for the data.  However, decision trees easily overfit data, and can be sensitive to small changes in the data.  They also tend to work best on fewer classes, so if training were performed on a larger set of cryptographic algorithms, it is possible that the model would break down.

To improve this model, further work could involve using a boosted decision tree, which Kolter and Maloof concluded was the best classifier for their work\cite{kolter}.%http://stats.stackexchange.com/questions/1292/what-is-the-weak-side-of-decision-trees

A model that produced comparable results on these experiments, but might generalize better to more algorithms and more realistic data is the SVM with a linear kernel.  Further work on optimizing the default scikit-learn parameters would be required.%More fine-tuning of the parameters would prevent over-fitting. %http://stats.stackexchange.com/questions/24437/advantages-and-disadvantages-of-svm

\section{Further Work}
In order to evaluate this on real-world examples, the test programs should be partitioned into basic blocks, then features extracted from those.  Then, the models should be evaluated on the feature set block by block.  This technique will extend this methodology to more general testing examples, and have the added benefit of identifying where in the program binary the crypto occurs, as well as finding multiple instances of it in a single binary.

Additionally, very simple loop features were extracted.  Possible improvements could be made by using the loop detection algorithm from Tubella and Gonz\`alez \cite{tubella} or LoopProf from Moseley, et. al\cite{Moseley}.

To increase the usefulness of these models, they should be trained on additional encryption and hashing algorithms, as well as examples from crypto libraries in addition to OpenSSL.  This should allow the models to detect a wider variety of cryptographic algorithms.

 % Results and Discussion

\lhead{}%\emph{Conclusion}}
\chapter{Conclusion}

This thesis examined the process of extracting features and training machine learning models for the detection and classification of cryptographic algorithms in compiled code.  Three different types of models were evaluated on four different feature sets using four different learning algorithms.  While the decision tree models were found to perform the best on these data, due to certain limitations of decision trees, it is possible that an SVM with a linear kernel will generalize better to real-world data.

Cross-validation results suggest that algorithm classification and detection will be over 95\% accurate, given a relatively small and homogeneous sample.  Furthermore, once this method has been implemented such that it examines the basic blocks of larger programs, it will be able to identify where in the binary program the crypto algorithm is executed, further simplifying the reverse engineer's task.

Further work is required to fully test these models in real world applications. However, this thesis demonstrates that it will be possible for machine learning algorithms to automatically detect and classify cryptographic primitives in binary code. This approach has the potential to outperform manual strategies in terms of both effectiveness and scalability. % Conclusion

%% ----------------------------------------------------------------
% Now begin the Appendices, including them as separate files

\addtocontents{toc}{\vspace{2em}} % Add a gap in the Contents, for aesthetics

\appendix % Cue to tell LaTeX that the following 'chapters' are Appendices

\addtocontents{toc}{\vspace{2em}}  % Add a gap in the Contents, for aesthetics
\backmatter

%% ----------------------------------------------------------------
\label{Bibliography}
\lhead{\emph{Bibliography}}  % Change the left side page header to "Bibliography"
\bibliographystyle{plain}  % Use the "unsrtnat" BibTeX style for formatting the Bibliography
\bibliography{bibliography}  % The references (bibliography) information are stored in the file named "bibliography.bib"

\end{document}